\documentclass{elsart}

\usepackage{graphicx}
\usepackage{amsfonts}

\begin{document}
\begin{frontmatter}

\title{Noncommutative geometry inspired charged black holes}
\author{Stefano Ansoldi}
  \thanks[udin]{email: ansoldi@trieste.infn.it}
  \address{Dipartimento di Matematica ed Informatica, Universit\`a  di
Udine, and Istituto Nazionale di Fisica Nucleare, Sezione di Trieste,
Trieste, Italy}
\author{Piero Nicolini}
\thanks[cmfd]{email: nicolini@cmfd.univ.trieste.it}
\address{Dipartimento di Matematica e Informatica, Consorzio di
Magnetofluidodinamica, Universit\`a degli Studi di Trieste and
Istituto Nazionale di Fisica Nucleare, Sezione di Trieste,
Trieste, Italy }

\author{Anais Smailagic}
 \thanks[smaila]{email: anais@ts.infn.it}
 \address{Istituto Nazionale di Fisica Nucleare, Sezione di Trieste,
 Trieste, Italy }

 \author{Euro Spallucci}
  \thanks[euro]{email: spallucci@trieste.infn.it}
  \address{Dipartimento di Fisica Teorica dell' Universit\`a  di
Trieste, and Istituto Nazionale di Fisica Nucleare, Sezione di Trieste,
Trieste, Italy}

\begin{abstract}
We find a new, non-commutative geometry inspired, solution of the
coupled Einstein-Maxwell field equations describing a
variety of charged, self-gravitating objects, including extremal
and non-extremal black holes. The metric smoothly interpolates
between  \textit{deSitter} geometry, at short distance, and
\textit{Reissner-Nordstr{\o}m } geometry far away from the origin.
Contrary to the ordinary Reissner-Nordstr{\o}m  spacetime there is
no curvature singularity in the origin neither ``naked'' nor
shielded by horizons.  We investigate both Hawking process and
pair creation in this new scenario.
  \end{abstract}
\end{frontmatter}

 In a recent paper we obtained a non-commutative geometry inspired solution
 of the Einstein equation smoothly interpolating between a deSitter core
 around the origin and an ordinary Schwarzschild spacetime at large
 distance \cite{nss}.
 The curvature singularity at $r=0$ is replaced by a regular deSitter vacuum
 accounting for the effect of non-commutative coordinate fluctuations at short
 distance.  Furthermore, the Hawking temperature does not blow up as the
 event horizon shrinks down, instead it reaches a maximum value for
 a radius $r_H\simeq 4.7\, \sqrt\theta$ and drops down to zero at
 $r_H\to r_0\simeq 3.0\, \sqrt\theta$; $\theta$ is the parameter
 measuring the amount of coordinate non-commutativity
 in the coordinate coherent states approach \cite{ncqft} :
$
\left[\, \mathbf{x}^\mu\ , \mathbf{x}^\nu\, \right]= i \,
\theta\, \epsilon^{\mu\nu}
$.
 Our model of ``noncommutative'' Schwarzschild black hole allows to
 answer the question: which is the endpoint of the Hawking process?
 This question has no definite answer neither in the framework of
 quantum field theory in curved spacetime, nor in string theory.
 In our case, we find a cold remnant given by
 a stable, extremal black hole, at zero temperature and radius $r_0$
 determined by the $\theta$ parameter.\\
  Our regular black hole has been employed as nonsingular background
geometry in the study of
 the vacuum polarization of matter by means of the effective
 action approach \cite{Spallucci:2006zj}.
 There are other
 applications of the coherent state approach to gravity
 \cite{Casadio:2005vg}, and recently our model has been extended to large
 extra-dimensions scenario in \cite{tom},
  where it is discussed under what conditions such a kind
 of remnant could be produced at LHC.
 With this exciting perspective
 in mind, we think it is phenomenologically relevant \footnote{$TeV$ black holes
 decay is presently an extremely important research field in view
 of the  experiments planned for the next generation particle colliders
 \cite{review}.}
 to extend our former
 solution to more general configurations carrying \textit{charges} other
 than pure gravitational one.  From this viewpoint,
 the immediate extension of the Schwarzschild-like objects we found
 in \cite{nss} amount to endow them with an ``Abelian hair''
 in the form of an $U\left(\,1\,\right)$ electric charge and  long
 range Coulomb-like electric field.  Objects of this sort will, eventually,
 decay, not only through Hawking radiation, but through pair creation
 of charged particles \cite{pairs}, as well, provided the near horizon electric
 field is higher than the threshold critical intensity $E_{cr}=\pi\, m_e^2/e$.
  In this case, the black hole can radiate either its whole charge or
  a fraction of it. In the former case it reaches a Schwarzschild phase,
  finally approaching the degenerate, extremal configuration discussed above.
  In the latter case, a stable  extremal Reissner-Nordstr{\o}m, with
  a residual charge, will be left out. \\
  In this letter we are going to solve Einstein-Maxwell equations in the
  presence of a static, spherically symmetric, minimal width, gaussian
  distribution of mass and charge. The decay of the black hole type solutions
  will be investigated both analytically and numerically along the lines
  presented above.\\
  Putting gravity aside for a moment,
 let us consider the effect of non-commuting coordinates on an elementary
 charge.
 We have shown that the coordinate fluctuations
  can be described, within coherent states approach, as a
 \textit{smearing effect} \cite{coherent}, turning point-like mass into
 a ``matter droplet'' with a gaussian profile. As coordinate
 non-commutativity is a property of the spacetime fabric itself, and not of
 its matter content, the same smearing effect is expected to operate on
 electric as well as ``inertial'' charges, i.e. masses. Thus, a point-charge
 $e$ is spread into a minimal width gaussian charge cloud according
 with the rule:

 \begin{equation}
 e\,\delta\left(\, \vec x\,\right)\longrightarrow
 \rho_{el.}\left(\,\vec x \,\right)=
\frac{e}{\left(\,4\pi\theta\,\right)^{3/2}}\,
\exp\left(-{\vec x}^2/4\theta\,\right)
 \end{equation}

 For a static, spherically symmetric charge distribution, the current density
  $J^\mu$ is  non-vanishing only along the time direction:

 \begin{equation}
 J^\mu\left(\,x\,\right)=
 \rho_{el.}\left(\, x \,\right)\,\delta^\mu_0
 \label{source}
 \end{equation}
The system of Einstein-Maxwell field equations reads
\begin{eqnarray}
&& R^\mu{}_\nu-\frac{1}{2}\, \delta^\mu{}_\nu\, R =
8\pi\,\left(\, T^\mu_\nu\vert_{matt.} + T^\mu_\nu\vert_{el.}  \,\right)
\label{einst}\\
&& \frac{1}{\sqrt{-g}}\, \partial_\mu\,\left(\, \sqrt{-g}\, F^{\mu\nu}\, \right)=
J^\nu \label{max}
\end{eqnarray}
where $T_{matt.}{}_\nu^\mu$ is the same as in \cite{nss}, while the
electromagnetic energy-momentum tensor is the usual one
with $F^{\mu\nu}=\delta^{0[\, \mu\,\vert}  \delta^{r\,\vert\, \nu \,]}\,
E\left(\, r\,\right) $.\\
The Coulomb-like field is obtained by solving the Maxwell equations (\ref{max})
with source (\ref{source}).
\begin{figure}[h!]
\begin{center}
\includegraphics[width=8cm,angle=0]{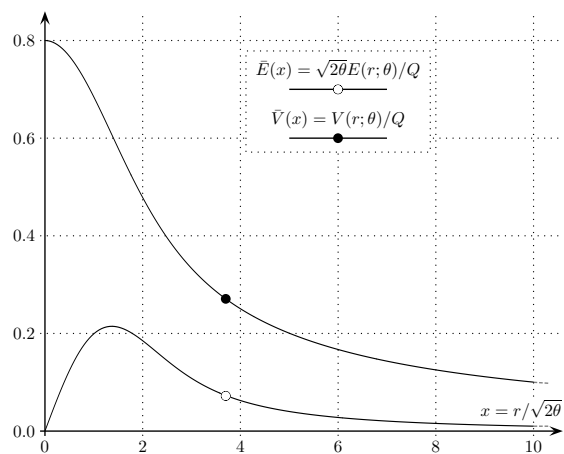}
\caption{\label{coulomb} \textit{
Electric field  $E\left(\, r\ ;\theta\,\right) $ (per unit charge)
and corresponding Coulomb-like potential $V\left(\, r\
;\theta\,\right) $ (per unit charge).
Smearing of the charge leads to a regular behavior at the origin.
The electric field reaches its maximum intensity near $r\approx 2\sqrt\theta$,
and then drops to zero. Long distance behavior is that of ordinary
Coulomb field.}}
\end{center}
\end{figure}

We find the electric field to be:

\begin{equation}
E\left(\ r\,\right)=\frac{2Q}{\sqrt\pi\, r^2}\,
\gamma\left(\, \frac{3}{2}\ ; \frac{r^2}{4\theta}\,\right)
\label{clmb}
\end{equation}

Then, solving the Einstein equations  (\ref{einst} ) we find
Reissner-Nordstr{\o}m like the metric :

\begin{eqnarray}
&& ds^2= -g_{00}\, dt^2 + g_{00}^{-1}\, dr^2 + r^2 d\Omega^2\label{ds}\\
&& g_{00}= 1
-\frac{2m\left(\, r\,\right)}{r} + \frac{Q^2}{\pi\, r^2}F\left(\, r\,\right)\ ,
\\
&&m\left(\, r\,\right)=\frac{2m_0}{\sqrt\pi}
\gamma\left(\, 3/2\ , r^2/4\theta\,\right)\ ,\\
&& F(r)\equiv  \gamma ^2\left(\,1/2\ ,r^2 /4\theta\,\right) -
 \frac{r}{\sqrt{2\theta }}\gamma\left(\, 1/2\ ,r^2 /2\theta\,\right)\ ,
 \label{rncbh}\\
&& \gamma\left(\, a/b\ ;x\,\right)\equiv
 \int_0^x \frac{du}{u}\, u^{a/b} \, e^{-u}
\end{eqnarray}

 \begin{figure}[h!]
\begin{center}
\includegraphics[width=15cm,angle=0]{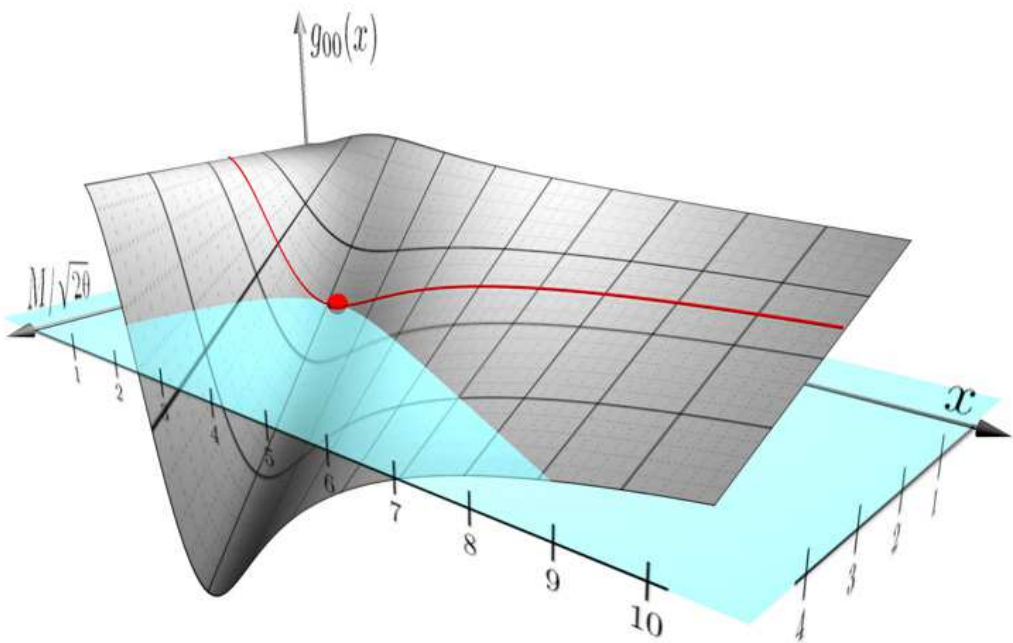}
\caption{\label{g00}
\textit{ $g_{00}$ versus $M$ and $r$, for a
charge, $Q=1$ in $\sqrt\theta$ units.  The intersection of the
  $g_{00}=0$ plane (cyan) with $g_{00}=g_{00}(r,M)$ surface (grey)
  gives the ``horizon curve'' whose minimum (red-dot) gives the extremal
  black hole. The portion of the surface below the plane $g_{00}=0$ represents
  the spacelike region between inner and outer horizons.}
}
\end{center}
\end{figure}

The line element (\ref{ds}) shows an  interesting asymptotic
behavior for small $r$. By using the asymptotic form of the lower incomplete
gamma-function one sees that $F\left(\, r\,\right)\sim O\left(\, r^6\,\right)$
and  the metric (\ref{ds})  is of the  deSitter type

\begin{equation}
 g_{00}= 1 -\frac{m_0}{3\sqrt{\pi}\, \theta^{3/2}}\, r^2
   +O\left(\, r^4\,\right)
 \label{desitter}
 \end{equation}
  The vacuum energy associated to the non-commuting coordinate fluctuations
 shows up as an effective cosmological
 $\Lambda_{eff.}=m_0/\sqrt{\pi}\, \theta^{3/2}$ leading to
 a \textit{finite
 curvature} in the origin: the ``singularity'' has been swept away by the
 vacuum fluctuation of the spacetime fabric itself. \footnote{
 A quite different type of regular, charged,
 black holes have been obtained, in the framework,
 of a new, non-linear, electrodynamics coupled to gravity \cite{bg}.}  \\
 It may seem surprising that there are no charge contributions to the
 effective cosmological constant. This is due to the linear behavior of the
 electric field at short distance, as seen from Fig.(\ref{coulomb}), which
 can only give contributions $O\left(\, r^4\,\right)$ to the metric. Thus,
 an observer close to the origin sees only the ``\textit{bare mass}'' $m_0$
 stripped from the charge dressing.\\
On the other hand, we want to compare our solution to the usual
Reissner-Nordstr{\o}m geometry, at large distance.
In this case, the asymptotic observer can only measure the
\textit{total mass-energy} in which he cannot distinguish anymore
between gravitation and electric contribution. Thus, the total mass-energy
is now define as the integrated  flux of $T_\mu^0$

\begin{equation}
M =\oint_\Sigma d\sigma^\mu\left(\,
T_\mu^0\vert_{matt.} + T_\mu^0\vert_{el.}  \,\right)
\label{mtot}
\end{equation}

where, $\Sigma$, is a $t=\mathrm{const.}$, closed three-surface.\\
 With the  definition  (\ref{mtot}) the metric (\ref{ds})
 reads

 \begin{equation}
  g_{00}= 1
-\frac{4M}{\sqrt\pi\, r} \gamma\left(\, 3/2\ , r^2/4\theta\,\right)
 + \frac{Q^2}{\pi\, r^2}\left[\, F\left(\, r\,\right)+\sqrt{\frac{2}{\theta}}
 \, r\, \gamma\left(\, 3/2\ , r^2/4\theta\,\right)\,\right]
 \label{ncrn}
  \end{equation}

The asymptotic form of (\ref{ncrn}) reproduces
Reissner-Nordstr{\o}m  metric for large distances.\\
Depending on  the values of $Q$ and $M$ metric  displays  different
causal structure: existence  of two horizons (~non-extremal black hole~),
\begin{figure}[h!]
\begin{center}
\includegraphics[width=15cm,angle=0]{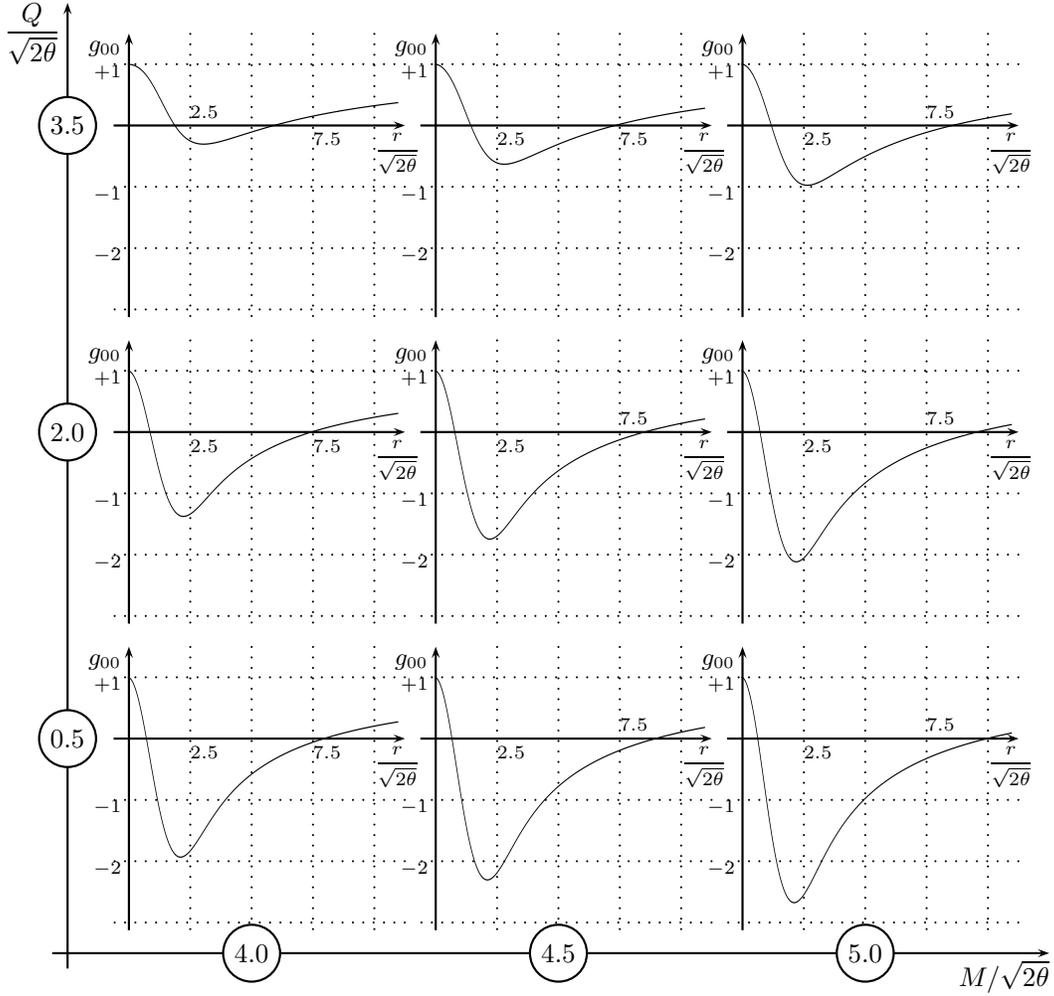}\caption{\label{nonextrbh}
\textit{$g_{00}$ as function of  $r$ for different
values of $M$ and $Q$, leading to two horizons. The intersections with the
$r$-axis gives inner and outer horizons. Given $M$ and increasing $Q$
$r_+ -r_- \to 0$ leading to extremal configuration. On the contrary, given $Q$
and increasing $M$,  enlarges the distance between horizons.}
}
\end{center}
\end{figure}
one horizon (~extremal black hole~)
 \begin{figure}[h!]
\begin{center}
\includegraphics[width=15cm,angle=0]{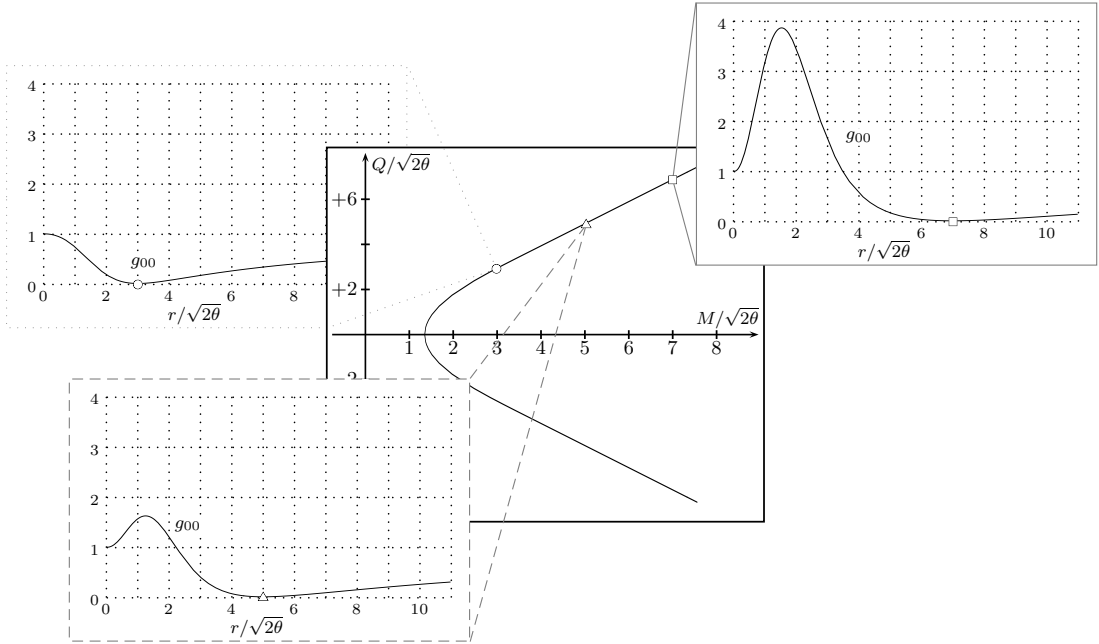}
\caption{\label{extrbh}
\textit{ $g_{00}$ as a function of $r$, for  $M$ and
$Q$ leading to a single, degenerate, horizon.
 Extremal horizons lie on the curve in the $Q, M$-plane (central plot) }
}
\end{center}
\end{figure}
 or no horizons (~massive charged ``droplet''~),
\begin{figure}[h!]
\begin{center}
\includegraphics[width=15cm,angle=0]{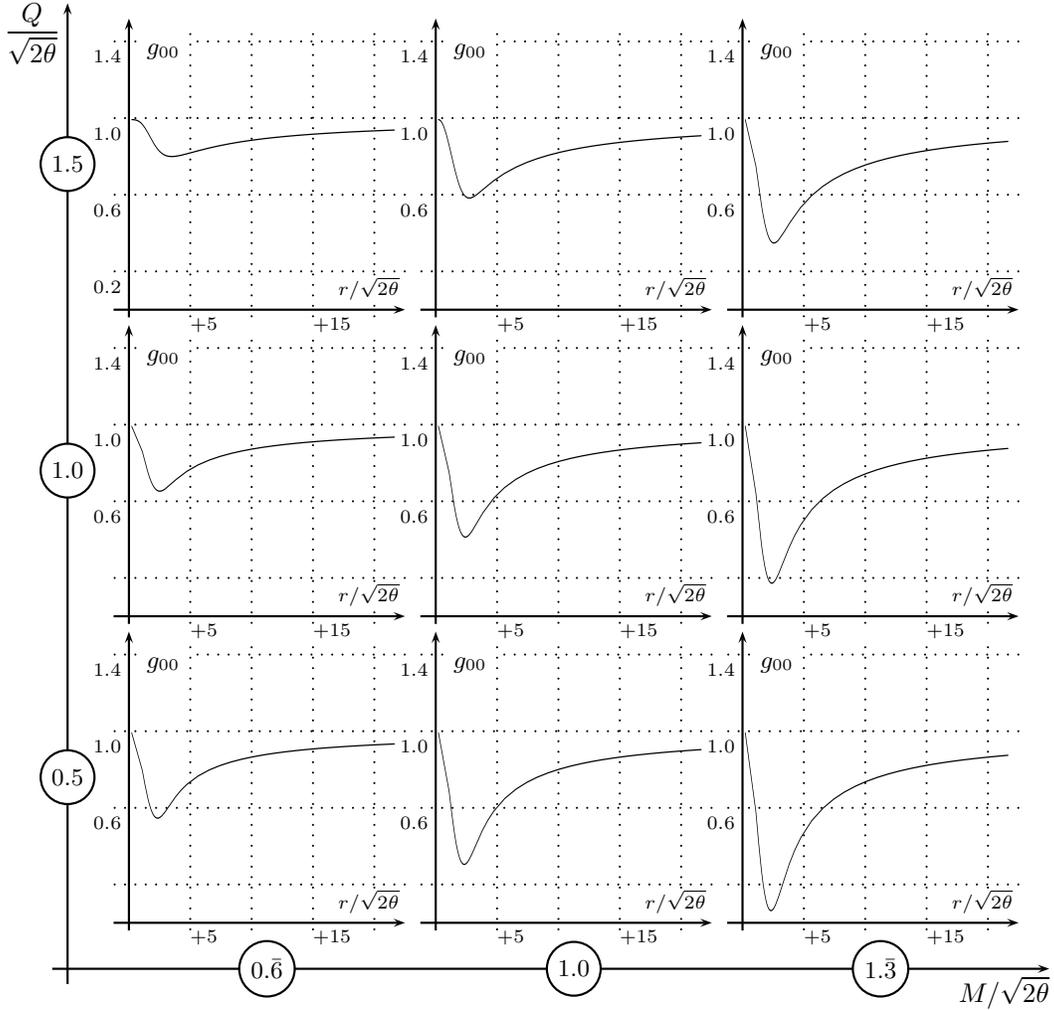}
\caption{\label{nobh}
\textit{$g_{00}$ as a function of $r$, for values of $M$
and $Q$ leading to no horizons. }
}
\end{center}
\end{figure}
 as can be seen from the plots of (\ref{ncrn}).
 $g_{00}\left(\, r_H\,\right)=0$
   cannot be solved analytically in $r_H$, in our case, but it allows to solve
   $M$ in terms of $r_H$  and $Q$ \footnote{ The use of the equation relating
  the total mass energy of the system to the radius of the event
  horizon follows the approach  proposed in
   \cite{bubbles}, with the advantage of allowing
   an indepeth investigation of geometry and dynamics of the system.}

   \begin{equation}
    M=\frac{Q^2}{2\sqrt{2\pi\theta}} +
  \frac{1}{\gamma\left(\, 3/2\ ,r^2_H /4\theta\,\right)}\left[\,
  \frac{\sqrt\pi}{4}r_H + \frac{Q^2}{4\sqrt\pi r_H }\,
  F\left(\,r_H \,\right)\,
  \right]\label{plot}
   \end{equation}
    Plotting (\ref{plot}) is an alternative way to investigate existence and
    location of event horizons.
    To understand the information that will follow from equation (\ref{plot})
    let us first study the simpler case of ordinary   Reissner-Nordstr{\o}m.
    The corresponding function $M\left(\, r_H\ ; Q\,\right)$ is
    $
    M=\left(\, r_H^2 +Q^2\,\right)/2r_H
    $.
     \begin{figure}[h!]
\begin{center}
\includegraphics[width=12cm,angle=0]{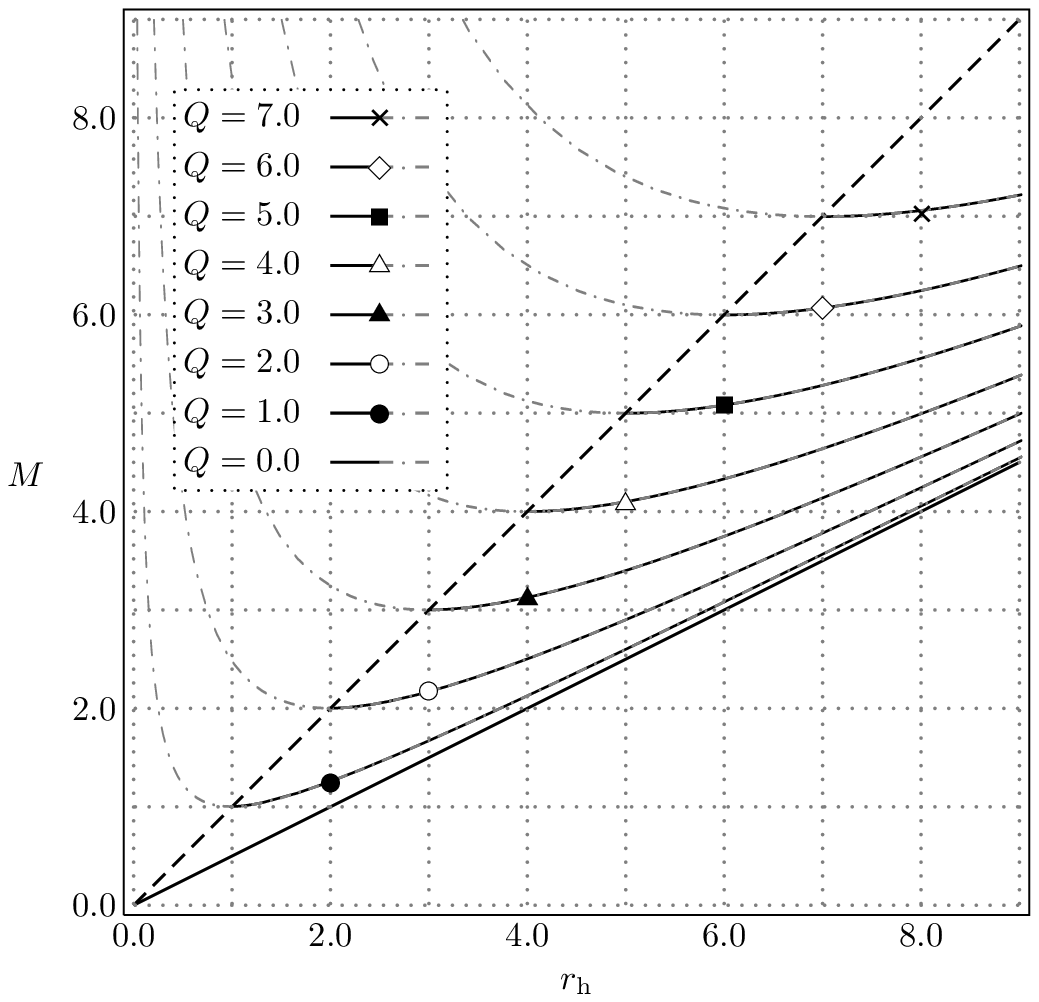}
\caption{\label{massa0}
\textit{   $M$ as a function of $r_H$ for different
$Q$ in case of Reissner-Nordstr{\o}m black hole.  The  straight line $M=r_H/2$
gives the Schwarzschild black-hole.
The dashed straight line $M=r_H$ intersects
the minima of $Q$-curves representing extremal Reissner-Nordstr{\o}m
black holes.  }
}
\end{center}
\end{figure}

\begin{figure}[h!]
\begin{center}
\includegraphics[width=12cm,angle=0]{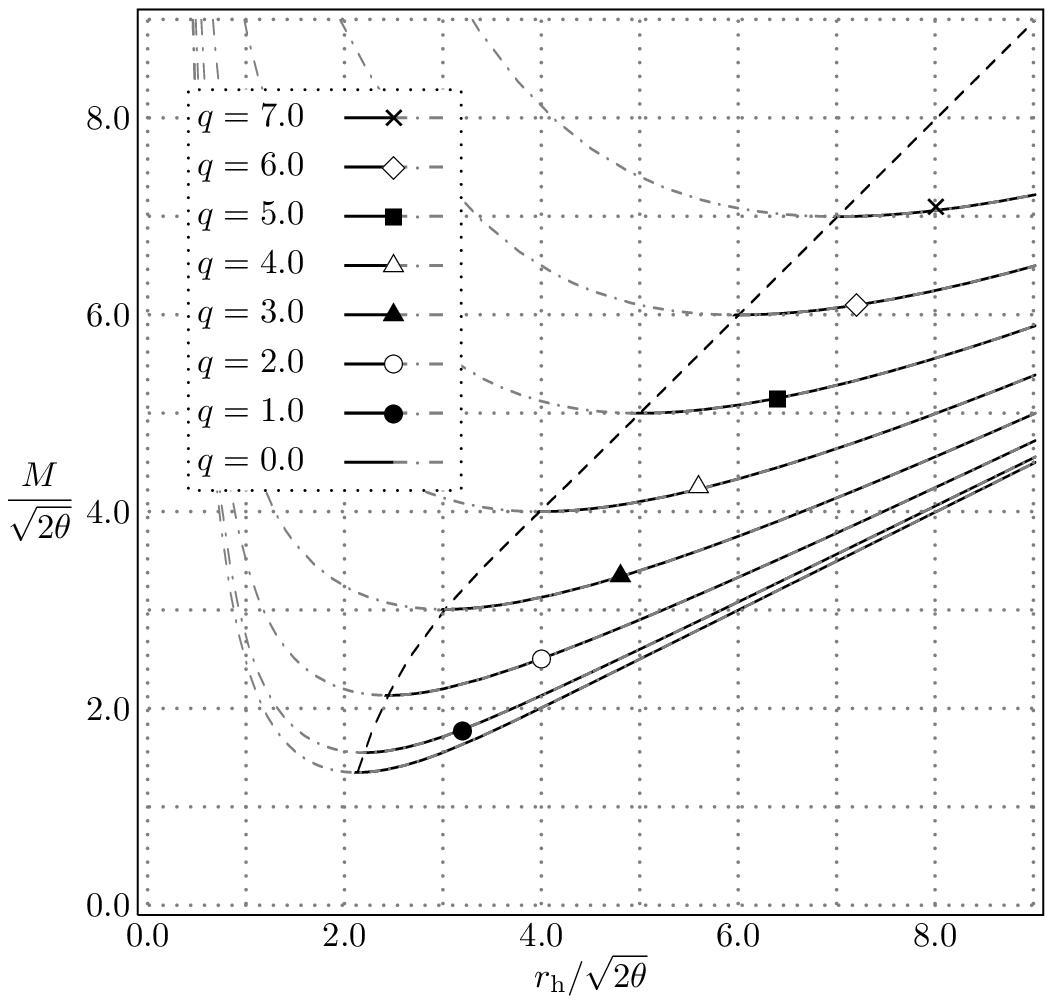}
\caption{\label{massa}
\textit{  $M$ as a function of $r_H$ for different
$Q$ in case of noncommutative Reissner-Nordstr{\o}m black hole.  The
Schwarzschild black-hole is given by the lowest curve, $Q=0$, which is no
more a straight line indicating that there can be two horizons for $M>
1.9\sqrt\theta$. Also in this case the dashed curve intersects the minima
of $Q$-curves representing extremal black holes. }
}
\end{center}
\end{figure}

\begin{figure}[h!]
\begin{center}
\includegraphics[width=12cm,angle=0]{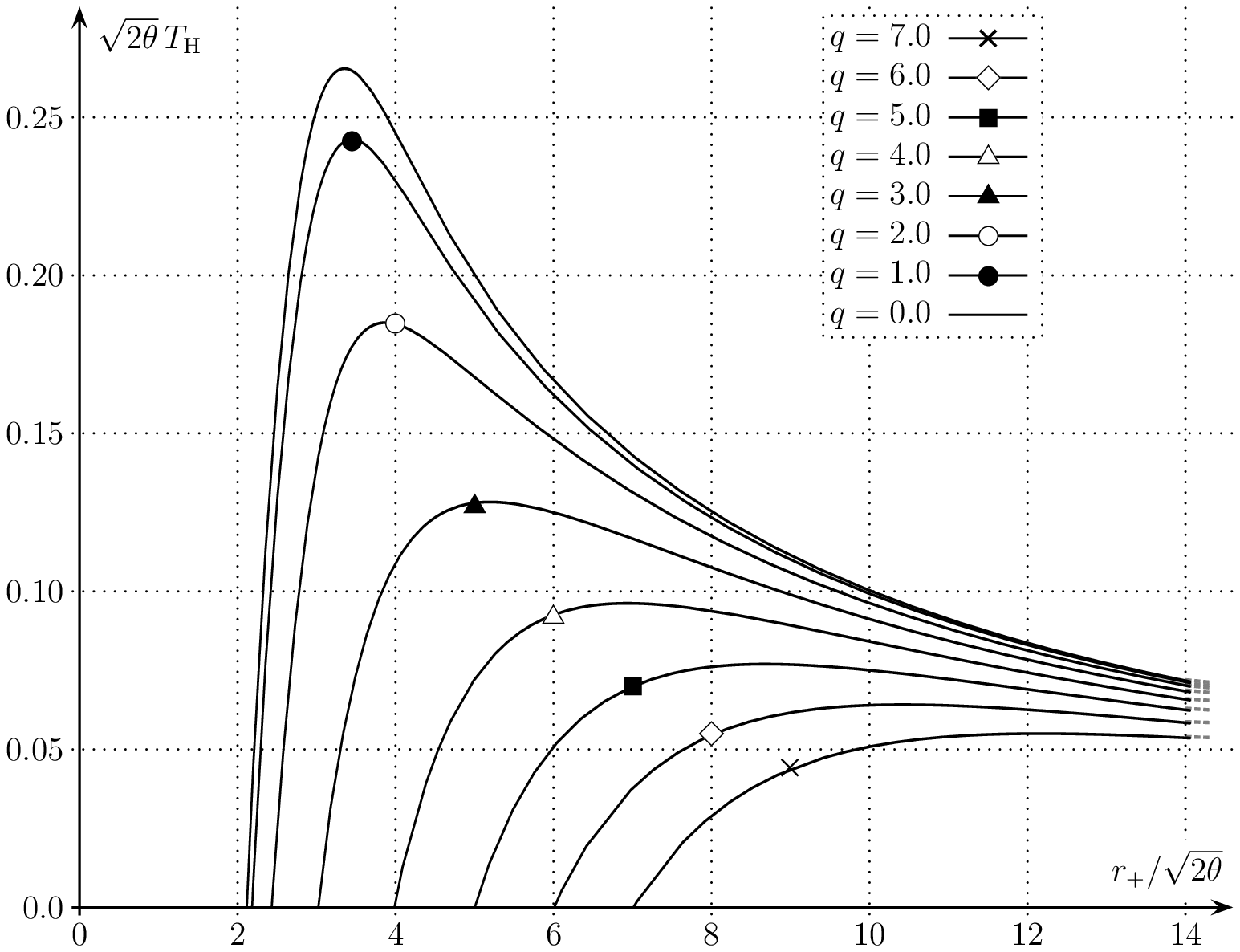}
\caption{\label{temp}
\textit{Hawking temperature as a function of
 $r_H$, for different values of $Q$.  The temperature drops to zero
 even in the case $Q=0$ as a result of coordinate uncertainty. The peak
 temperature drops with increasing $Q$.}
}
\end{center}
\end{figure}

  The  plot Fig.(\ref{massa0}) consists of a  family of
  curves labelled by the value of the electric charge $Q$.
  The horizons, and the corresponding values of $M$, are read as the
  intersections of the grid lines,  with each curve in Fig.(\ref{massa0}).
  It turns out that the minima
  of the curves define the extremal black hole configurations. All degenerate
  horizons lie on the dashed straight line $M=r_H$. The full straight
  line   $M=r_H/2$   corresponds to the  $Q=0$ Schwarzschild black hole.
  Fig.(\ref{massa}) indicates that, for any given charge $Q$, the total mass
  energy of the system is \textit{minimized} by  the extremal black hole
  configuration.\\
  Thus, we conclude that every Reissner-Nordstr{\o}m black  hole has natural
  tendency to evolve towards its extremal configuration. \\
  The same kind of analysis can now be performed in the non-commutative
  case. In spite of complicated looking functions, e.g. $M=M(R_H;Q)$  ,
  we reach  essentially same conclusions as discussed above.  For this
  purpose we plotted (\ref{plot}) in Fig.(\ref{massa}).
 Once  again we see that the \textit{extremal curve} connects the minima of
 $Q$-curves, and
  asymptotically approaches the line $M=r_H$,  in agreement with  the behavior
  of the metric (\ref{ds}) at large distance. The difference between
  Fig.(\ref{massa0}) and Fig.(\ref{massa}) is seen at short distance where
  we find a non-vanishing
  \textit{minimal mass}, for $Q=0$, corresponding to the extremal
  Schwarzschild-like black hole found in \cite{nss}.
  This effect is due to the non-commutativity of spacetime coordinates.
  \\
   Fig.(\ref{massa}) suggests the following scenario of the dynamical evolution
   of noncommutative inspired Reissner-Nordstr{\o}m like black holes.
  At the very beginning system is described by
  $Q$ and $r_+$, then it starts evolving towards its ground
  state, i.e. tends to minimize its total mass-energy. It can do it in two
  different ways: i) it can radiate away its mass
  mainly in form of neutral particles ( Hawking evaporation ),
  ``sliding down'' the $Q$-curve ; ii) it can
  reduce both its mass and charge through pair creation, jumping
  from higher to lower $Q$-curves.
  In the former case the final configuration would be  an extremal
  Reissner-Nordstr{\o}m-like  black hole
  with a residual charge, while in the latter it ends up as
   an extremal Schwarzschild black hole. Which evolution path
   will be taken, essentially depends
  on the interplay between two decay channels.\\
  In connection with the above picture, it is important to investigate
  the behavior of the  Hawking temperature $T_H$.
  The temperature $T_H$ is given by:
\begin{eqnarray}
 4\pi \, T_H &&= \frac{1}{r_+}
 \left[ {1 - \frac{{r^3_+ \exp ( - r^2_+ /4\theta )}}{{4\theta ^{3/2}
 \gamma\left(\,3/2\ , r^2_+ /4\theta\,\right)}}} \right]+ \nonumber\\
 &&-\frac{{4Q^2 }}{{\pi r^3_+ }}
 \left[ {\gamma ^2\left(\,3/2\ ,r^2_+ /4\theta \,\right) +
 \frac{{r^3_+ \exp ( - r^2_+ /4\theta )}}
 {{16\,\theta ^{3/2} \gamma\left(\,3/2\ , r^2_+ /4\theta\,\right)}}
 F(r_+)} \right]
\end{eqnarray}
We plot $T_H$ in Fig.(\ref{temp}).
From Fig.(\ref{temp}) and Fig.(\ref{massa}) one infers that,
instead of growing indefinitely temperature reaches a maximum
value and then drops to zero at the extremal black hole
configuration. This is the same behavior already encountered in
the noncommutative neutral case \cite{nss}.
The effect of charge is just to lower the maximum temperature.\\
At this point we would like to discuss, at least qualitatively, which
of the possible decay channels is more likely. \\
The creation of $e^\pm$ pairs near the event horizon is
 described by the Schwinger formula \cite{dunne}. This formula implies
 that in order for creation process to take place, the electric field
 has to exceed the \textit{critical intensity} $E_{cr.}=\pi \, m^2_e/e$. In
 our case this condition leads to

 \begin{equation}
 \frac{2Q}{\sqrt\pi\, r^2_+}\,
\gamma\left(\, \frac{3}{2}\ ; \frac{r^2_+}{4\theta}\,\right)
\ge \pi\, \frac{m^2_e}{e} \label{schw}
\end{equation}

From  Fig.[\ref{coulomb}] one sees that the electric field attains
maximal intensity near $r\approx 2\sqrt\theta$, for any $Q$:
$ E_{max.}=E\left(\, r_{max.}\,\right)\approx 0.28\, Q\,\sqrt{2\theta} $.\\
 Let us express the total
charge $Q$ as an integer multiple of the fundamental charge $e$,
i.e. $Q=ze$. Equation (\ref{schw}), with any possible value for
$r_+$, implies that there is creation for any $z\ge 1$ because
$z\propto m_e^2 \theta << 1$.

\begin{figure}[h!]
\begin{center}
\includegraphics[width=15cm,angle=0]{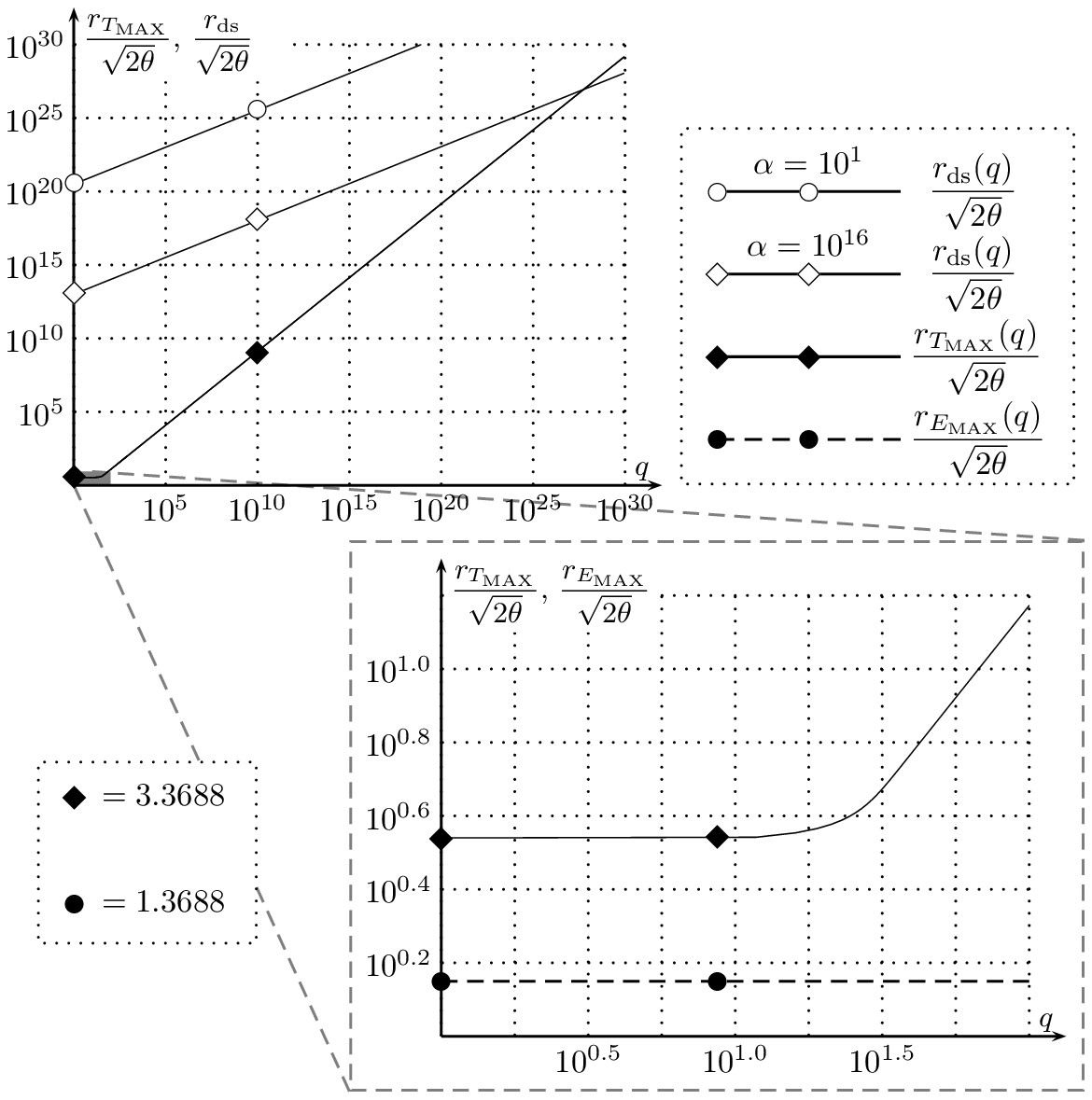}
\caption{\label{diado}\textit{Plot of the dimensionless radius at which the
temperature
of the black hole attains its maximum value,
$r _{T _{\mathrm{MAX}}}/\sqrt{2\theta}$, as
a function of the dimensionless charge $q = Q / \sqrt{2 \theta}$. All diagram
are in a
bi-logarithmic scale; moreover the dimensionless charge $q$ is expressed as
a multiple of
the elementary charge (for instance, on the $q$ axis, the value $0$ corresponds to
the electron charge, the value $10 ^{2}$ to the charge of $100$ electrons,
and so on).
In the top diagram the dimensionless radius of the dyadosphere
$r _{\mathrm{ds}}/\sqrt{2\theta}$
is also plotted for comparison (the ratio between
the non commutativity scale, $\sqrt{2 \theta}$, and the Planck length,
$l _{\mathrm{P}}$,
is called $\alpha$, i.e. $\sqrt{2 \theta} = \alpha l _{\mathrm{P}}$).
We see that for a
wide range of charges and for values of $\alpha$ between $10$ and $10 ^{16}$,
$r _{\mathrm{ds}}/\sqrt{2\theta}$
largely exceeds $r _{T _{\mathrm{MAX}}}/\sqrt{2\theta}$.
In the bottom diagram the details of the above plot for a number of elementary
charges between
$1$ and $10 ^{2}$ is shown. In this case, we also plot the dimensionless radius
$r _{E _{\mathrm{MAX}}}/\sqrt{2\theta}$ at which the electric fields
reaches its maximum value (dashed line): this shows that always
$r _{E _{\mathrm{MAX}}}/\sqrt{2\theta} < r _{T _{\mathrm{MAX}}}/\sqrt{2\theta}$
and also that
$r _{E _{\mathrm{MAX}}}/\sqrt{2\theta} < r _{{\mathrm{ds}}}/\sqrt{2\theta}$. }
}
\end{center}
\end{figure}
Thus, we conclude that the electric field at the horizon
is \textit{always strong enough}
to create pairs.
In fact, over-criticality could be attained
 even far away from  $r_+$ . To verify this possibility
 we use the concept of \textit{dyadosphere} introduced
in \cite{ruffini}. Dyadosphere represents  the spherical region,
of radius $r_{ds}$, where $E$ is strong enough to produce pairs.
In our case the radius $r_{ds}$ is determined by

\begin{equation}
\frac{r_{ds}^2}{\gamma\left(\,\frac{3}{2}\ ;\frac{r^2_{ds}}{4\theta}\,\right)}=
\frac{2z e^2}{\pi^{3/2} m^2_e}
r_{ds}^2\simeq \frac{2z e^2}{\pi^{3/2} m^2_e}\,
\gamma\left(\,\frac{3}{2}\ ;\frac{z e^2  }{4\pi \theta m^2_e }\,\right)
\label{rdyado}
\end{equation}

As a first approximation we find

\begin{equation}
r_{ds}^2\simeq \frac{2z e^2}{\pi^{3/2} m^2_e}\,
\gamma\left(\,\frac{3}{2}\ ;\frac{z e^2  }{4\pi \theta m^2_e }\,\right)
\label{dyado}
\end{equation}

Equation (\ref{dyado}) show that the characteristic length scale
of the dyadosphere is the electron Compton wavelength,  which is
orders of magnitude larger than $\sqrt\theta$. Thus, we conclude
that, for charged black holes, described by the line element
(\ref{rncbh}), $r_{ds} \gg r_+$. As a consequence, these black holes
are extremely unstable under pair creation. This analysis implies
that Schwinger mechanism should dominate the early phase of black
hole decay. Hawking radiation follows Schwinger phase. At the end,
one is left with a neutral massive degenerate remnant  found in
\cite{nss}.

In this letter we have investigated the existence of charged black
holes inspired by non-commutativity of the space time manifold at
short distance. We have found a Reissner-Nordstr{\o}m like metric.
It reproduces exactly ordinary  Reissner-Nordstr{\o}m solution at
large distance, while at short distance it gives deSitter
spacetime in place of a curvature singularity. Furthermore we have
investigated the fate of such black hole against quantum decay. We
came to the conclusion, that among possible decay channels,
Schwinger pair production dominates the early stage while Hawking
radiation determines the final stage. Recently, it has been
proposed that our black holes could be produced at LHC \cite{tom}.
This exciting phenomenological perspective could provide an
experimental verification of our models.


\ack
We would like to thank Dr. Thomas G. Rizzo for useful discussions
about the subject of this paper.

    \end{document}